\documentclass[10pt]{iopart}
\usepackage{graphicx}
\usepackage{amssymb}
\eqnobysec

\usepackage{iopams}
\usepackage[colorlinks,linkcolor=phthaloblue,urlcolor=black,citecolor=phthaloblue]{hyperref}
\usepackage{xcolor}
\usepackage{color}
\definecolor{phthaloblue}{rgb}{0.0, 0.06, 0.54}

\newcommand{\beq}{\begin{equation}}
\newcommand{\eeq}{\end{equation}}
\newcommand{\beqa}{\begin{eqnarray}}
\newcommand{\eeqa}{\end{eqnarray}}
\newcommand{\cc}{{\cal C}}
\newcommand{\dd}{{\rm d}}
\newcommand{\erf}{\mathop{\rm erf}}
\newcommand{\frad}[2]{{\displaystyle{\displaystyle#1\over\displaystyle#2}}}
\newcommand{\mean}[1]{\langle#1\rangle}
\newcommand{\prob}{\mathbb{P}}
\newcommand{\w}[1]{{\tilde{#1}}}
\newcommand{\F}[1]{{\widehat{#1}}}

\begin{document}

\title{Survival probability of random walks and L\'evy flights with stochastic resetting}

\author{Claude Godr\`eche and Jean-Marc Luck}

\address{Universit\'e Paris-Saclay, CNRS, CEA, Institut de Physique Th\'eorique,
91191~Gif-sur-Yvette, France}

\begin{abstract}
We perform a thorough analysis of the survival probability of symmetric random
walks with stochastic resetting,
defined as the probability for the walker not to cross the origin
up to time $n$.
For continuous symmetric distributions of step lengths
with either finite (random walks) or infinite variance (L\'evy flights),
this probability can be expressed
in terms of the survival probability of the walk without
resetting, given by Sparre Andersen theory.
It is therefore universal, i.e., independent of the step length distribution.
We analyze this survival probability at depth,
deriving both exact results at finite times and asymptotic late-time results.
We also investigate the case where the step length distribution is
symmetric but not continuous,
focussing our attention onto arithmetic distributions
generating random walks on the lattice of integers.
We investigate in detail the example of the simple Polya walk
and propose an algebraic approach for lattice walks with a larger range.
\end{abstract}

\address{\today}

\eads{\mailto{claude.godreche@ipht.fr}, \mailto{jean-marc.luck@ipht.fr}}

\maketitle

\section{Introduction}
\label{intro}

This work is a sequel to the recent study of the statistics of the maximum and of the
number of records for random walks with stochastic resetting~\cite{us}, a topic
of much current interest (see~\cite{emsrev} for a review).
Consider a random walk starting from the origin, defined by the recursion
\beq
x_{n+1}=\left\{
\matrix{
0\hfill &\hbox{with prob.~$r$},\hfill\cr
x_n+\eta_{n+1}\quad &\hbox{with prob.~$1-r$}.
}
\right.
\label{def}
\eeq
At each time step, the walker is reset to the origin with probability $r$.
The step lengths~$\eta_n$ are iid (independent and identically distributed)
random variables with an arbitrary symmetric distribution with density $\rho(\eta)$.

The central object of interest of the present work is the survival probability
(or persistence probability) of the random walker,
that is, the probability for the walker
not to cross the origin, up to time~$n$:
\beq
Q_n=\prob(x_1\ge0,\dots,x_n\ge0)=\prob(x_1\leq0,\dots,x_n\leq0).
\label{sdef}
\eeq

The generating series of the sequence $Q_n$ has the expression
\beq
\w Q(z)=\sum_{n\ge0}Q_nz^n=\frac{\w q((1-r)z)}{1-rz\,\w q((1-r)z)},
\label{sresintro}
\eeq
where $\w q(z)$ is the generating series of the sequence of survival probabilities $q_n$
of the same walker in the absence of resetting.
The identity~(\ref{sresintro}) can be deduced either
from a renewal approach (see~section~\ref{renew}),
or from the formalism used
in~\cite{us} for investigating the statistics of the maximum.
The expression of $\w q(z)$ stems from Sparre Andersen theory (see~section~\ref{sa}).
If the step length distribution is continuous and symmetric,
with either finite (random walks)
or infinite variance (L\'evy flights),~$\w q(z)$ is given by~(\ref{qser}),
and so $q_n$ and $Q_n$ are universal:
$q_n$ only depends on $n$, whereas $Q_n$ also depends on the resetting probability $r$.
If the step length distribution possesses a discrete component,~$\w q(z)$ is
given by the general expression~(\ref{gqser}) and is no longer universal.

The relation~(\ref{sresintro}) is the starting point of a thorough analysis
of several novel aspects of the problem, as we now summarize.
Section~\ref{sa} is a brief reminder of the predictions of Sparre Andersen theory
on the survival probability $q_n$ of symmetric random walks in the absence of resetting,
the step length distribution being either continuous or not.
In section~\ref{renew}, using a direct renewal approach,
we establish in full generality the identity~(\ref{sresintro}) between the
survival probabilities of the random walk in the presence and in the absence of resetting.
We also recall how the same result can be deduced from the study made in~\cite{us}.
The case of a symmetric continuous step length distribution is considered in section~\ref{cont}.
We analyze the universal expression of the survival probability $Q_n$ at depth,
deriving both exact results at finite times and asymptotic late-time results.
Section~\ref{polya} is devoted to the simple Polya walk on the lattice of integers,
for which the survival probability is studied following the same setting.
More general arithmetic step length distributions are dealt with in
section~\ref{arith} by means of a novel algebraic approach.

\section{Elements of Sparre Andersen theory}
\label{sa}

This section is a brief reminder of some aspects of Sparre Andersen theory
on sums of random variables~\cite{sparre53,sparre54,feller2}.
Consider a random walk without resetting,
whose position at time~$n$,
\beq
x_n=\eta_1+\cdots+\eta_n,
\eeq
is the sum of $n$ iid
random variables whose common distribution is symmetric, with density $\rho(\eta)$.
Introduce the probabilities
\beqa
\pi_n&=&\prob(x_n\ge0)=\prob(x_n\leq0),
\nonumber\\
q_n&=&\prob(x_1\ge0,\dots,x_n\ge0)=\prob(x_1\le0,\dots,x_n\le0).
\eeqa
A combinatorial theorem due to Sparre Andersen states that these probabilities
are related by the identity~\cite[ch.~XII]{feller2}
\beq
\w q(z)=\sum_{n\ge0}q_nz^n=\exp\Bigg(\sum_{n\ge1}\frac{\pi_n}{n}\,z^n\Bigg).
\label{saiden}
\eeq
The first few of these relations read
\beq
q_0=1,\quad
q_1=\pi_1,\quad
q_2=\frac{\pi_2+\pi_1^2}{2},\quad
q_3=\frac{2\pi_3+3\pi_1\pi_2+\pi_1^3}{6}.
\eeq
Let us now discuss the consequences of the identity~(\ref{saiden}),
depending on the nature of the distribution~$\rho(\eta)$.

\subsubsection*{Continuous symmetric distributions.}

If the step length distribution is symmetric and continuous,
the position $x_n$ is non-zero with certainty at all times $n\ge1$,
so that we have
\beq
\pi_n=\frac{1}{2}\qquad(n\ge1).
\label{pdemi}
\eeq
The series entering the right-hand side of~(\ref{saiden}) therefore reads
\beq
\w q(z)=\sum_{n\ge0}q_nz^n=\frac{1}{\sqrt{1-z}},
\label{qser}
\eeq
hence
\beq
q_n=b_n,
\label{qn}
\eeq
where $b_n$ denotes the binomial probability
\beq
b_n=\frac{(2n)!}{(2^nn!)^2}=\frac{{2n\choose n}}{2^{2n}}.
\label{bdef}
\eeq
This is the well-known universal expression of the survival probability
for a symmetric continuous distribution,
with either finite or infinite variance~\cite{sparre53,sparre54,feller2}.
The following power-law decay and asymptotic series of corrections are
therefore universal:
\beq
q_n=\frac{1}{\sqrt{\pi
n}}\left(1-\frac{1}{8n}+\frac{1}{128n^2}+\frac{5}{1024n^3}+\cdots\right).
\label{qasy}
\eeq

\subsubsection*{Arbitrary symmetric distributions.}

In the general case where the step length distri\-bution is symmetric,
but not necessarily continuous,
its density $\rho(\eta)$ may contain Dirac delta functions,
either at the origin $(\eta=0)$ or at pairs of symmetric positions $(\eta=\pm a)$.
In such a circumstance, the probability
\beq
P_n=\prob(x_n=0)
\label{Pdef}
\eeq
is non-zero for some $n$.
Equations~(\ref{pdemi}) and~(\ref{qser}) become
\beq
2\pi_n=\prob(x_n\ge0)+\prob(x_n\le0)=1+P_n
\eeq
and
\beq
\w
q(z)=\sum_{n\ge0}q_nz^n
=\frac{1}{\sqrt{1-z}}\exp\Bigg(\frac12\sum_{n\ge1}\frac{P_n}{n}\,z^n\Bigg).
\label{gqser}
\eeq
We have in particular
\beq
\w q(z)\approx\frac{E}{\sqrt{1-z}}\qquad(z\to1),
\eeq
hence
\beq
q_n\approx\frac{E}{\sqrt{\pi n}},
\eeq
where
\beq
E=\exp\Bigg(\frac12\sum_{n\ge1}\frac{P_n}{n}\Bigg)\ge1
\label{Edef}
\eeq
is the asymptotic enhancement factor of the survival probability $q_n$
with respect to the formula~(\ref{qasy}),
which holds for symmetric continuous distributions
(see~\cite[ch.~XVIII]{feller2}, and~\cite{us,mms}).

The special class of arithmetic step length distributions, that is,
discrete probability distributions yielding
random walks on the lattice of integers,
will be considered in section~\ref{arith}.

\section{A direct renewal approach}
\label{renew}

\begin{figure}[!ht]
\begin{center}
\includegraphics[angle=0,width=.5\linewidth,clip=true]{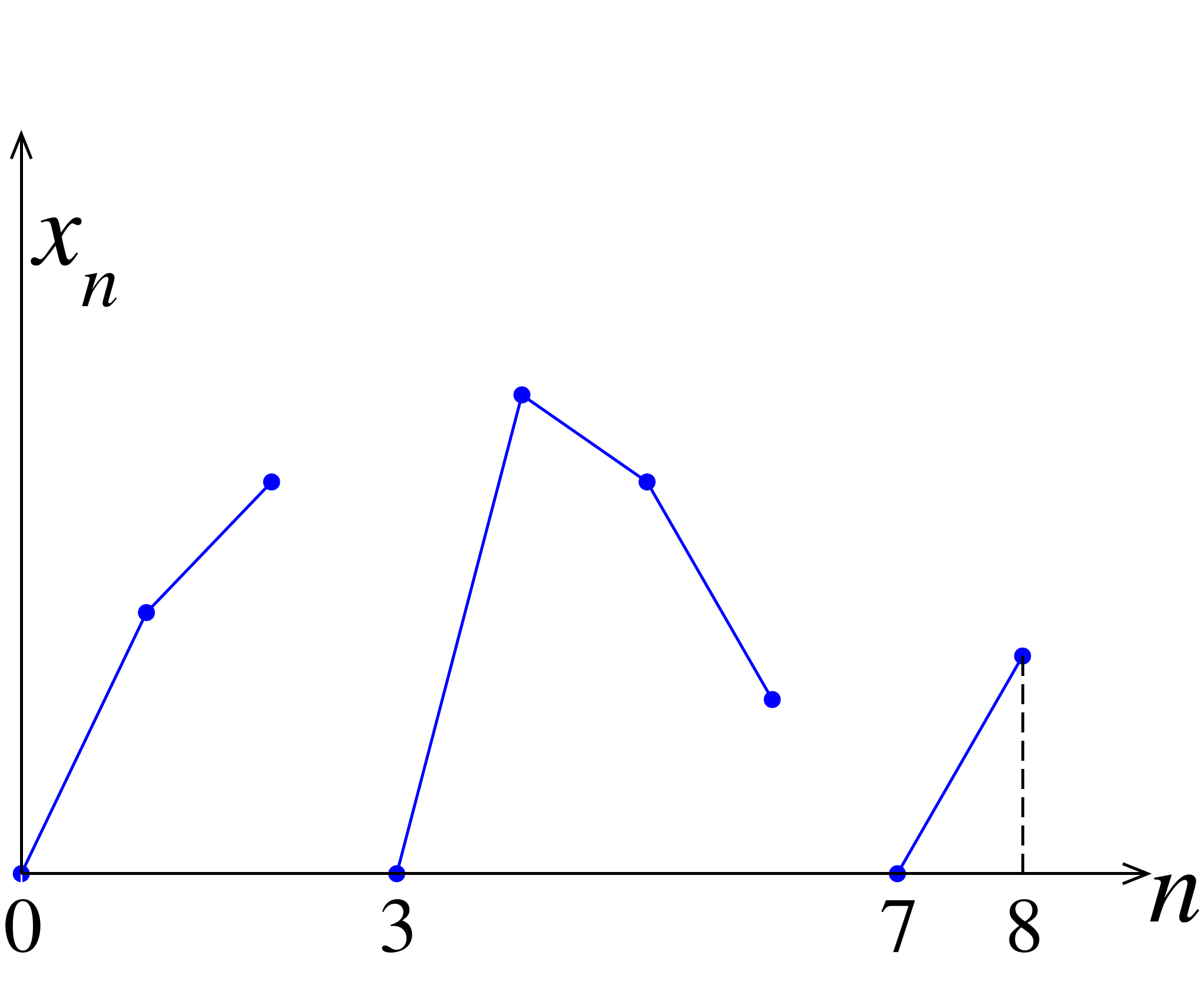}
\caption{\small
Example of a random walk of $n=8$ steps which does not cross the origin,
with two resetting events at times 3 and 7.
In terms of a renewal process (see text),
we have $N_8=2$, $T_1=3$, $T_2=4$ and $B_8=1$.}
\label{walk}
\end{center}
\end{figure}

The sequence of resetting events is a usual renewal process on the integers.
The description of such a process is a simple transcription
of that given for renewal processes with continuous random variables (see, e.g.,~\cite{glrenew}).
The durations $T_1,T_2,\dots$ between successive resetting events
are iid with common geometric distribution
\beq
f_k=\prob(T=k)=r(1-r)^{k-1}\qquad(k\ge1),
\label{fdef}
\eeq
whose complementary distribution function reads
\beq
g_k=\prob(T> k)=\sum_{j> k}f_j=(1-r)^k\qquad(k\ge0).
\eeq
For any given time $n\ge0$, the number of resetting events up to $n$ is
the unique integer $N_n\ge0$ such that
\beq
T_1+\cdots+T_{N_n}\le n<T_1+\cdots+T_{N_n+1},
\label{nsum}
\eeq
and the age $B_n$ of the process at time $n$,
also dubbed the backward recurrence time, is such that
\beq
n=T_1+\cdots+T_{N_n}+B_n,\qquad B_n=0,1,\dots,T_{N_n+1}-1.
\label{bndef}
\eeq
These definitions are illustrated in figure~\ref{walk}.

Let us now relate
the survival probability $Q_n$ of the walker in the presence of resetting
to the survival probability $q_n$ of the same walker in the absence of resetting.
For a given number of steps $n$, let us denote a realization of the process by
\beq\label{eq:cc}
\cc=\{N_n=m,T_1=k_1,\dots,T_{m}=k_{m},B_n=b\}.
\eeq
The joint probability distribution associated to~(\ref{eq:cc}) reads
\beq
P(\cc)
=f_{k_1}\ldots f_{k_m}\, g_b\;\delta\Big(\sum_{i=1}^m k_i+b,n\Big),
\eeq
where $\delta$ is the Kronecker delta symbol.

In the example shown in figure~\ref{walk}, $\cc=\{N_8=2,T_1=3,T_2=4,B_8=1\}$,
which gives a contribution $q_2q_3q_1$ to $Q_8$.
For a generic realization $\cc$, this contribution reads
\beq
Q(\cc)=q_{k_1-1}\dots q_{k_{m}-1}\,q_{b},
\label{sprod}
\eeq
because every duration $T_i=k_i$ between two consecutive resetting events
brings a factor $q_{k_i-1}$,
since the walker survives during $k_i-1$ steps before a resetting occurs.
The last incomplete lapse of time $B_n=b$ brings a factor $q_b$.

The survival probability $Q_n$ is obtained by averaging~(\ref{sprod})
over all realizations $\cc$:
\beqa
Q_n&=\sum_{\cc}P(\cc)Q(\cc)
\\
&=\sum_{m\ge0}\sum_{k_1,\dots,k_m}\sum_{b}f_{k_1}\dots f_{k_m}g_b
\;q_{k_1-1}\dots q_{k_{m}-1}q_b\,\delta\Big(\sum_{i=1}^mk_i+b,n\Big).
\nonumber
\eeqa
The associated generating series follows easily:
\beqa
\w Q(z)&=
\sum_{m\ge0}\sum_{k_1,\dots,k_m}\sum_{b}f_{k_1}z^{k_1}\dots f(k_m)z^{k_m}g_bz^b
\;q_{k_1-1}\dots q_{k_{m}-1}q_{b}
\nonumber\\
&=\beta(z)\sum_{m\ge0}\alpha(z)^m=\frac{\beta(z)}{1-\alpha(z)},
\eeqa
with
\beqa
\alpha(z)=\sum_{k\ge1}f_k\,q_{k-1}\,z^k=rz\,\w q((1-r)z),
\nonumber
\\
\beta(z)=\sum_{b\ge0}g_b\,q_b\,z^b=\w q((1-r)z).
\eeqa

We thus obtain the key result
\beq
\w Q(z)=\frac{\w q((1-r)z)}{1-rz\,\w q((1-r)z)},
\label{sres}
\eeq
relating the generating series of $\w Q(z)$ and $\w q(z)$,
already announced in~(\ref{sresintro}).

As the resetting probability $r$ varies,
the expression~(\ref{sres}) interpolates between the following two limiting situations.
In the absence of resetting ($r=0$), we have $\w Q(z)=\w q(z)$,
and so $Q_n=q_n$ for all $n$.
In the other limit ($r=1$),
where the walker stays put at the origin,~(\ref{sres}) yields $\w Q(z)=1/(1-z)$,
and so $Q_n=1$ for all $n$, as expected.

The first few relations between the $Q_n$ and the $q_n$ read
\beqa
Q_1&=&(1-r)q_1+r,
\nonumber\\
Q_2&=&(1-r)^2q_2+2r(1-r)q_1+r^2,
\nonumber\\
Q_3&=&(1-r)^3q_3+r(1-r)^2(q_1^2+2q_2)+3r^2(1-r)q_1+r^3.
\eeqa

There is an alternative route to obtaining~(\ref{sres}), based upon
the relationship between the survival probability $Q_n$
and the distribution of the maximum of the walk,
\beq
M_n=\max(0,\dots,x_n).
\eeq
All surviving walks on the negative side up to time $n$,
obeying $x_1\leq0,\dots,x_n\leq0$,
have $M_n=0$,
whereas all non-surviving walks have $M_n>0$.
We have therefore
\beq
Q_n=\prob(M_n=0).
\eeq
The generating series of this quantity, obtained by means of a renewal integral equation,
identifies to~(\ref{sres})~\cite{us}.

To close, let us mention that rational expressions similar to~(\ref{sres})
relating generating series for quantities with and without stochastic resetting
are in fact ubiquitous~\cite{us,emsrev,em,paf,mmss,m2s2}.
The above direct renewal approach applies to all these situations.

\section{Symmetric continuous step length distributions}
\label{cont}

\subsection{Generating series}
\label{contgal}

In this section we consider the case of symmetric continuous step length distributions,
with either finite or infinite variance.
The survival probability $q_n$ in the absence of resetting
is given by the universal formula~(\ref{qn}),
the generating series $\w q(z)$ of which is~(\ref{qser}).
Inserting this latter expression into~(\ref{sres}) yields
the following generating series of the survival probabilities $Q_n$
for symmetric continuous step length distributions:
\beq
\w Q(z)=\frac{1}{\sqrt{1-(1-r)z}-rz}.
\label{sercont}
\eeq
This universal expression
is one of the cornerstones of the present work.
We now investigate its consequences in detail.

\subsection{Exact finite-time results}
\label{contex}

\begin{figure}[!ht]
\begin{center}
\includegraphics[angle=0,width=.7\linewidth,clip=true]{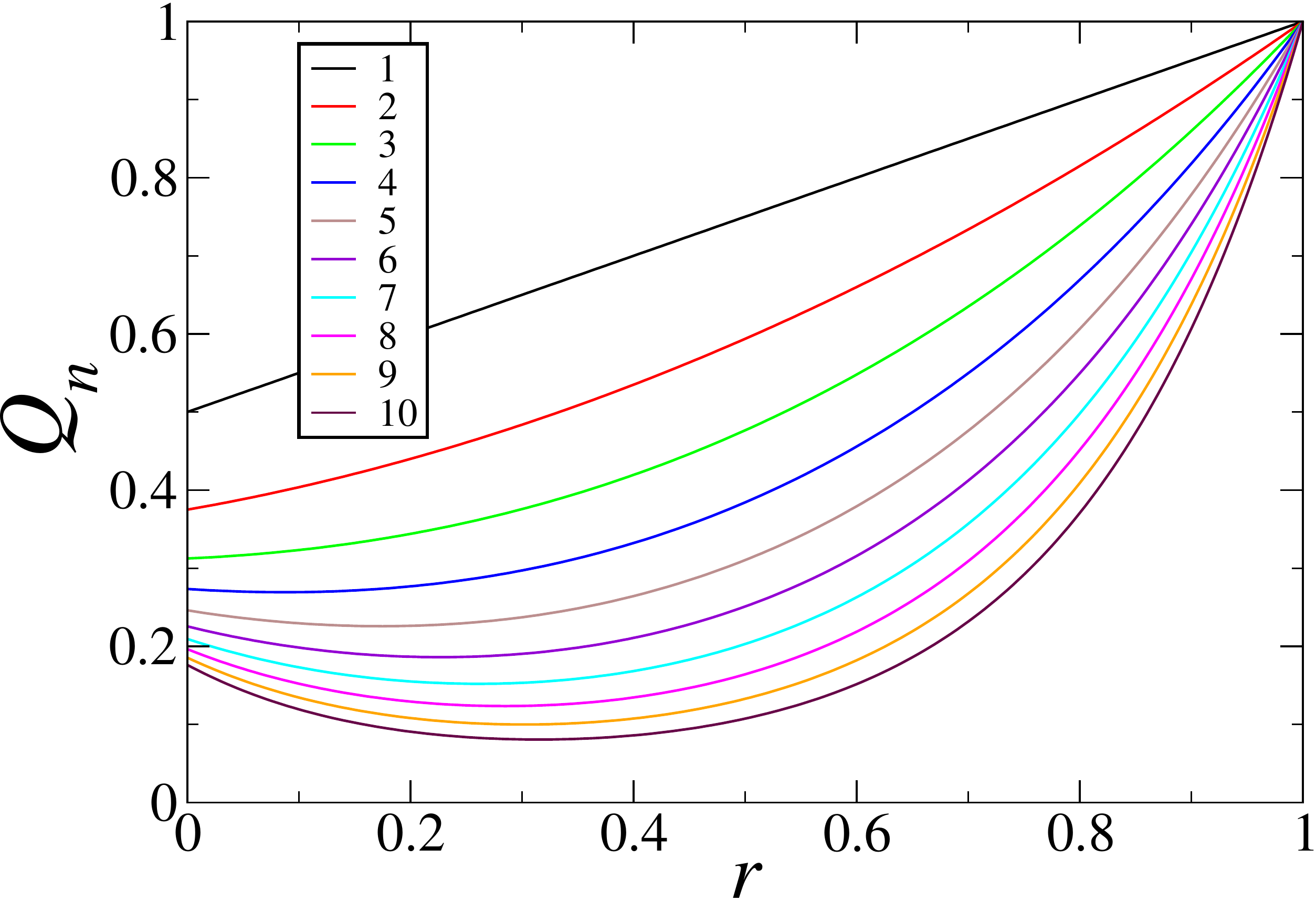}
\caption{\small
Survival probability $Q_n$ for symmetric continuous step length distributions
against resetting probability $r$ for times $n$ up to 10 (see legend).}
\label{qcont}
\end{center}
\end{figure}

The survival probability $Q_n$
only depends on time $n$ and on the resetting probability~$r$.
By expanding the formula~(\ref{sercont}) as a power series in $z$,
it is readily seen that $Q_n$ is a polynomial of degree~$n$ in $r$,
which has the same universality as the expression~(\ref{qn})
to which it reduces for $r=0$.
It is plotted against~$r$ in figure~\ref{qcont} up to $n=10$.
Its first values read
\beqa
Q_0=1,
\nonumber\\
Q_1=\frac{1}{2}(1+r),
\nonumber\\
Q_2=\frac{1}{8}(3+2r+3r^2),
\nonumber\\
Q_3=\frac{1}{16}(5+r+7r^2+3r^3),
\nonumber\\
Q_4=\frac{1}{128}(35-12r+66r^2+20r^3+19r^4),
\nonumber\\
Q_5=\frac{1}{256}(63-59r+166r^2-6r^3+75r^4+17r^5).
\eeqa

It turns out that $Q_n$ obeys a four-term linear recursion,
whose origin can be traced back to the works by Abel on algebraic functions
(see~\cite{bostan} for an overview including historical and algorithmic
aspects).
In the present situation, the derivation goes as follows.
First, by eliminating the square root in~(\ref{sercont}),
we obtain a quadratic equation for $\w Q(z)$:
\beq
(1-(1-r)z-r^2z^2)\w Q(z)^2-2rz\w Q(z)-1=0.
\label{algcont}
\eeq
Second, by differentiating the above equation w.r.t.~$z$
and judiciously eliminating non-linear terms,
we obtain a linear first-order differential equation for $\w Q(z)$:
\beqa
&&2(1-(1-r)z)(1-(1-r)z-r^2z^2)\w Q'(z)
\nonumber\\
&&+(r-1+(1+r)(1-3r)z+3r^2(1-r)z^2)\w Q(z)
\nonumber\\
&&-r(2-(1-r)z)=0.
\eeqa
Third, by expanding the above differential equation as a power series in $z$,
we obtain the four-term linear recursion
\beqa
&&2n Q_n-(4n-3)(1-r)Q_{n-1}+((2n-3)(1-2r)-3r^2)Q_{n-2}
\nonumber\\
&&+(2n-3)r^2(1-r)Q_{n-3}=2r\delta_{n1}-r(1-r)\delta_{n2}.
\label{recfour}
\eeqa

The behavior of the survival probability $Q_n$ as $r\to0$ or $r\to1$
can be studied by appropriately expanding the generating
series~(\ref{sercont}).
In the weak-resetting regime ($r\to0$),
we obtain
\beq
\w
Q(z)=\frac{1}{\sqrt{1-z}}+\left(\frac{z}{1-z}-\frac{z}{2(1-z)^{3/2}}\right)r+\cdots,
\eeq
hence
\beq
Q_n=b_n+c_n r+\cdots,\qquad c_n=1-nb_n.
\label{cont0}
\eeq

\begin{figure}[!ht]
\begin{center}
\includegraphics[angle=0,width=.7\linewidth,clip=true]{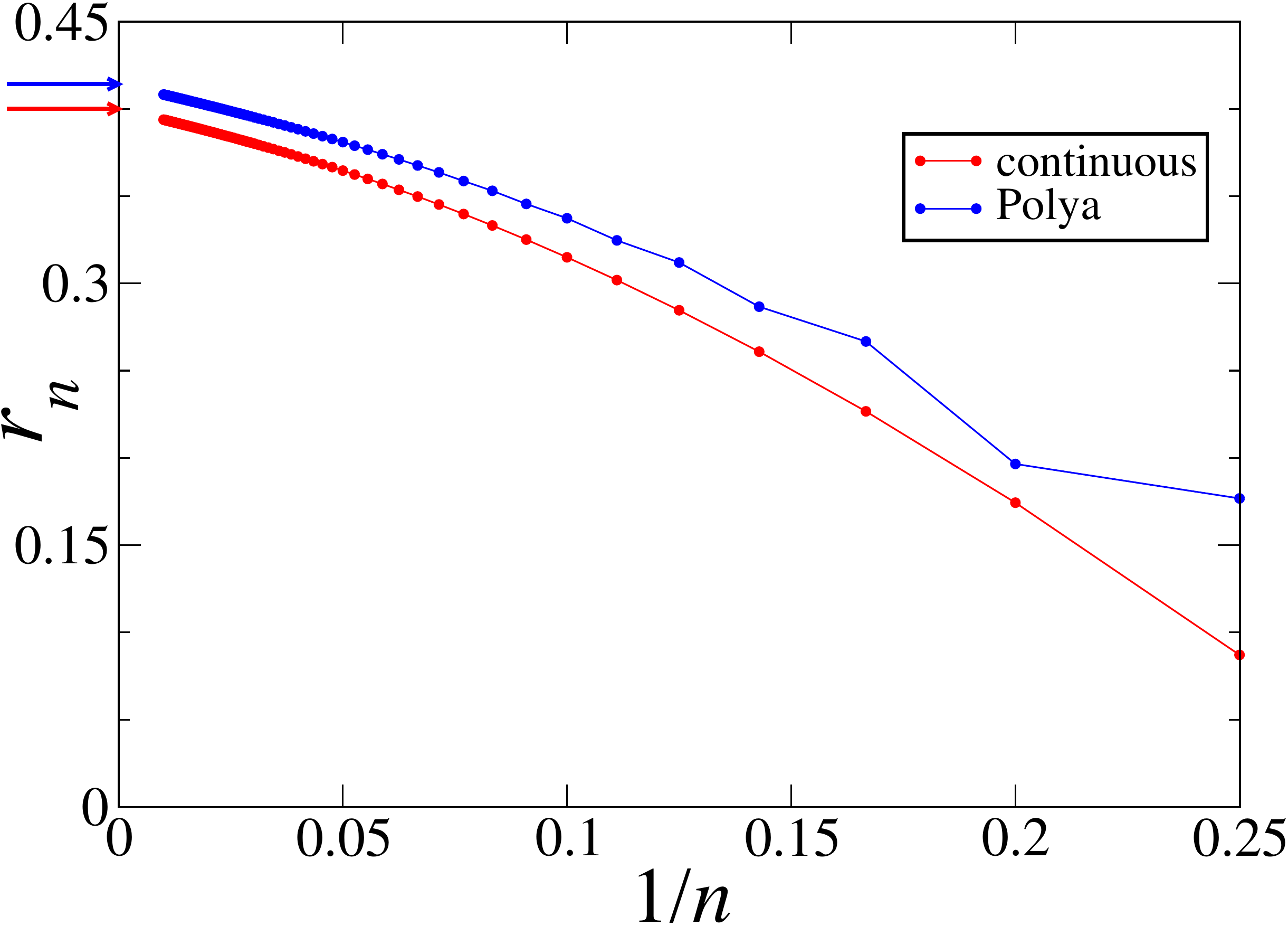}
\caption{\small
Values $r_n$ of the resetting probability
at which $Q_n$ is minimal against $1/n$ for $n\ge4$.
Red: data for for symmetric continuous step length distributions.
Blue: data for the simple Polya walk (see section~\ref{polyaex}).
Arrows: limits~(\ref{rmaxcont}) and~(\ref{rmaxpolya}).}
\label{rn}
\end{center}
\end{figure}

The first three correction terms $c_1=1/2$, $c_2=1/4$, $c_3=1/16$ are positive,
whereas $c_4=-3/32$ and all subsequent ones are negative.
This corroborates the observation (see figure~\ref{qcont})
that $Q_n$ is monotonically increasing with $r$ for $n=1$, 2 and~3,
whereas it exhibits a minimum for a non-trivial $r_n$ for all $n\ge4$.
The value $r_n$ of the resetting probability
at which $Q_n$ is minimal is plotted against $1/n$ in figure~\ref{rn} (red dataset).
For large times, $r_n$ approaches the limit~(\ref{rmaxcont}) (red arrow),
at which the asymptotic decay rate $K$~(\ref{kcont}) is maximal.

In the strong-resetting regime ($r\to1$),
we obtain
\beq
\w Q(z)=\frac{1}{1-z}-\frac{z}{2(1-z)^2}(1-r)+\cdots,
\eeq
hence
\beq
Q_n=1-\frac{n}{2}\,(1-r)+\cdots
\label{cont1}
\eeq
When $r=1$, as said above, the walker stays put at the origin, so that $Q_n=1$ at all times.
The correction term expresses
that there are on average $n(1-r)$ time steps where the walker is not being reset,
and the walker's displacement at each of those times is positive (or negative)
with probability $\pi_1=1/2$.

\subsection{Asymptotic late-time results}
\label{contasy}

Let us now turn to the asymptotic behavior of the survival probabilities $Q_n$ at late times.
For any fixed value of the resetting probability $r$,
the closest singularity of the generating series~(\ref{sercont})
is a simple pole located at
\beq
z_0=\frac{\sqrt{5r^2-2r+1}+r-1}{2r^2},
\eeq
whereas the branch-point singularity of the square root lies further away, at
\beq
z_c=\frac{1}{1-r}.
\eeq
We have indeed $1<z_0<z_c$.
Therefore $Q_n$ decays exponentially as
\beq
Q_n\approx A\,\e^{-Kn},
\label{expcont}
\eeq
with
\beq
K=\ln z_0=\ln\frac{\sqrt{5r^2-2r+1}+r-1}{2r^2}.
\label{kcont}
\eeq
The amplitude $A$ can also be worked out and reads
\beq
A=\frac{2r}{\sqrt{5r^2-2r+1}}.
\eeq

An exponential decay law of the survival probability was to be expected.
In the presence of resetting,
the walker indeed reaches a nonequilibrium steady state
characterized by a non-trivial stationary distribution of its position~\cite{emsrev}.
As is well known, for stationarity processes,
the survival probability (or persistence probability)
generically falls off exponentially in time.
This picture is corroborated by the fact that the number of factors in~(\ref{sprod})
scales linearly with time.

\begin{figure}[!ht]
\begin{center}
\includegraphics[angle=0,width=.7\linewidth,clip=true]{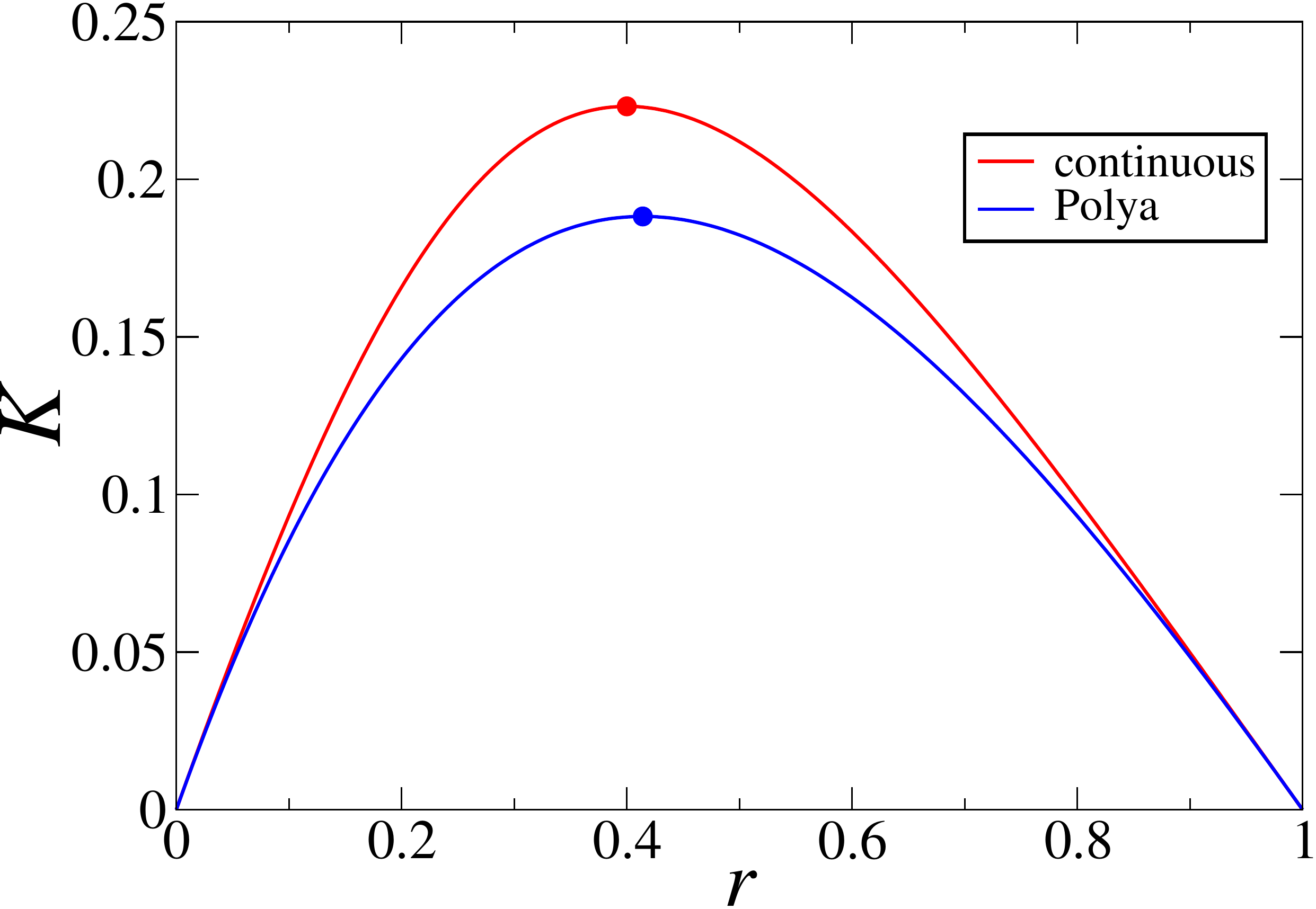}
\caption{\small
Asymptotic decay rate $K$ of the survival probabilities
against resetting probability $r$.
Red: universal expression~(\ref{kcont})
for symmetric continuous step length distributions.
Blue: expression~(\ref{kpolya}) for the simple Polya walk.
Symbols: maximal values $K_{\rm max}$
(see~(\ref{kmaxcont}),~(\ref{rmaxcont}) and~(\ref{kmaxpolya}),~(\ref{rmaxpolya})).}
\label{kplot}
\end{center}
\end{figure}

The decay rate $K$~(\ref{kcont}) entering the exponential law~(\ref{expcont})
is plotted against~$r$ in figure~\ref{kplot} (red curve).
It vanishes as $r\to0$ and $r\to1$, according to
\beq
\matrix{
K=r-\frad{r^2}{2}+\cdots\hfill & (r\to0),\cr
K=\frad{1-r}{2}-\frad{7(1-r)^3}{48}+\cdots\quad & (r\to1),
}
\eeq
and reaches its maximum,
\beq
K_{\rm max}=\ln\frac{5}{4}\approx0.223143,
\label{kmaxcont}
\eeq
for the value
\beq
r=\frac{2}{5}
\label{rmaxcont}
\eeq
of the resetting probability.

Finally, a crossover takes place in the scaling regime where $n$ is large, whereas $r$ is small.
Both singularities of the generating series~(\ref{sercont}) merge in this regime,
as we have $K\approx z_0-1\approx r$, whereas $z_c-z_0\approx r^2$ is much smaller.
Setting $z=\e^{-s}$,
the leading-order singular behavior of~(\ref{sercont}) reads
\beq
\w Q(z)\approx\frac{1}{\sqrt{s+r}},
\eeq
hence
\beq
Q_n\approx\frac{\e^{-rn}}{\sqrt{\pi n}}.
\label{scacont}
\eeq
The power-law decay~(\ref{qasy}) of the survival probability $q_n$ in the
absence of resetting
is thus reduced by an exponential factor involving the mean number of
resettings $nr$.
The full crossover between the decay laws~(\ref{expcont}) and~(\ref{scacont})
is captured by the more complete scaling formula
\beq
\w Q(z)\approx\frac{1}{\sqrt{s+r}-r}\approx\frac{\sqrt{s+r}+r}{s+r(1-r)},
\eeq
featuring a pole at $s=-r(1-r)$ and a square-root branch point at $s=-r$,
and yielding
\beq
Q_n\approx\e^{-rn}\left(\frac{1}{\sqrt{\pi
n}}+r\,\e^{r^2n}\left(1+\erf(r\sqrt{n})\right)\right).
\eeq
The first and second term of this crossover formula
respectively match the asymptotic decay
of the expressions~(\ref{scacont}) and~(\ref{expcont}),
to leading order at small $r$.
The crossover time between both regimes diverges as $n\sim1/r^2$.

\section{The simple Polya walk}
\label{polya}

\subsection{Generating series}
\label{polyagal}

The simple Polya walk on the lattice of integers is generated by the binary
step length distribution
\beq
\rho(\eta)=\frac{1}{2}\left(\delta(\eta-1)+\delta(\eta+1)\right).
\label{rhopol}
\eeq
In this case, the probability $P_n$ introduced in~(\ref{Pdef}) can be
determined as follows.
In the absence of resetting,
the walker's position $x_n$ is zero if the number of steps $n=2k$ is even,
and the walk consists of $k$ steps to the right and $k$ steps to the left.
This reads (see~(\ref{bdef}) for the definition of $b_k$)
\beq
P_{2k}=b_k,\qquad P_{2k+1}=0.
\label{Ppolya}
\eeq
We have thus, using~(\ref{qser}),
\beq
\sum_{k\ge1}\frac{b_k}{k}\,y^k=\int_0^y\left(\frac{1}{\sqrt{1-x}}-1\right)\frac{\dd x}{x}
=-2\ln\frac{1+\sqrt{1-y}}{2}.
\eeq
The sum involved in~(\ref{gqser}) therefore reads
\beqa
\frac{1}{2}\sum_{n\ge1}\frac{P_n}{n}\,z^n
&=&\frac{1}{4}\sum_{k\ge1}\frac{b_k}{k}\,z^{2k}
\nonumber\\
&=&-\frac{1}{2}\ln\frac{1+\sqrt{1-z^2}}{2}
\nonumber\\
&=&\ln\frac{\sqrt{1+z}-\sqrt{1-z}}{z}.
\eeqa
We are thus left with the expressions
\beq
\w q(z)
=\frac{2}{\sqrt{1-z^2}+1-z}
=\frac{1}{z}\left(\frac{1+z}{\sqrt{1-z^2}}-1\right).
\label{qserpolya}
\eeq
The second formula is more amenable to power-series expansion,
as it shows explicitly the even and odd components of $\w q(z)$.

The survival probability $q_n$ without resetting
inherits the dependence on the parity of $n$ of the probability $P_n$ given
in~(\ref{Ppolya}).
It reads
\beq
q_{2k-1}=q_{2k}=b_k,
\label{qnpolya}
\eeq
and obeys the power-law fall-off
\beq
q_n\approx\sqrt\frac{2}{\pi n}.
\label{qasypolya}
\eeq
The enhancement factor with respect to the universal result~(\ref{qasy})
therefore reads
\beq
E=\sqrt2.
\label{epolya}
\eeq
The corrections to the leading behavior~(\ref{qasypolya}) depend on the parity of $n$,
according~to
\beq
q_n=\sqrt\frac{2}{\pi n}\,\left\{\matrix{
\left(1-\frad{1}{4n}+\frad{1}{32n^2}+\cdots\right)\hfill & (n\hbox{ even}),\hfill
\cr
\left(1-\frad{3}{4n}+\frad{25}{32n^2}+\cdots\right)\hfill & (n\hbox{ odd}).\hfill
}
\right.
\label{qasyserpolya}
\eeq

Finally, inserting~(\ref{qserpolya}) into~(\ref{sres}), we obtain
\beqa
\w Q(z)
&=&\frac{2}{\sqrt{1-(1-r)^2z^2}+1-(1+r)z}
\nonumber\\
&=&1+\frac{(1+r)z+(1+r^2)z^2}{\sqrt{1-(1-r)^2z^2}+1-(1+r^2)z^2}.
\label{serpolya}
\eeqa
These expressions
are another key outcome of this work.
We now investigate their consequences in detail.

\subsection{Exact finite-time results}
\label{polyaex}

The survival probability $Q_n$ of the Polya walk with resetting
only depends on time $n$ and on the resetting probability~$r$.
By expanding the second line of~(\ref{serpolya}) as a power series in $z$,
we obtain $Q_0=1$, as should be, and
\beq
Q_{2k+1}=(1+r)R_k,\qquad Q_{2k+2}=(1+r^2)R_k\qquad(k\ge0),
\label{QR}
\eeq
where the generating series of the $R_k$ reads, upon setting $y=z^2$,
\beq
\w R(y)=\sum_{k\ge0}R_ky^k=\frac{1}{\sqrt{1-(1-r)^2y}+1-(1+r^2)y}.
\label{serR}
\eeq
As it turns out,
$R_k$ is a reciprocal polynomial of degree $2k$ in $r$,
and so $Q_n$ is a polynomial of degree~$n$ in $r$,
whose structure depends on the parity of $n$, according to~(\ref{QR}).
For $r=0$, we have $R_k=b_{k+1}$, in agreement with~(\ref{qnpolya}).
For $r=1$, we have $R_k=1/2$, and so $Q_n=1$, as expected.

\begin{figure}[!ht]
\begin{center}
\includegraphics[angle=0,width=.7\linewidth,clip=true]{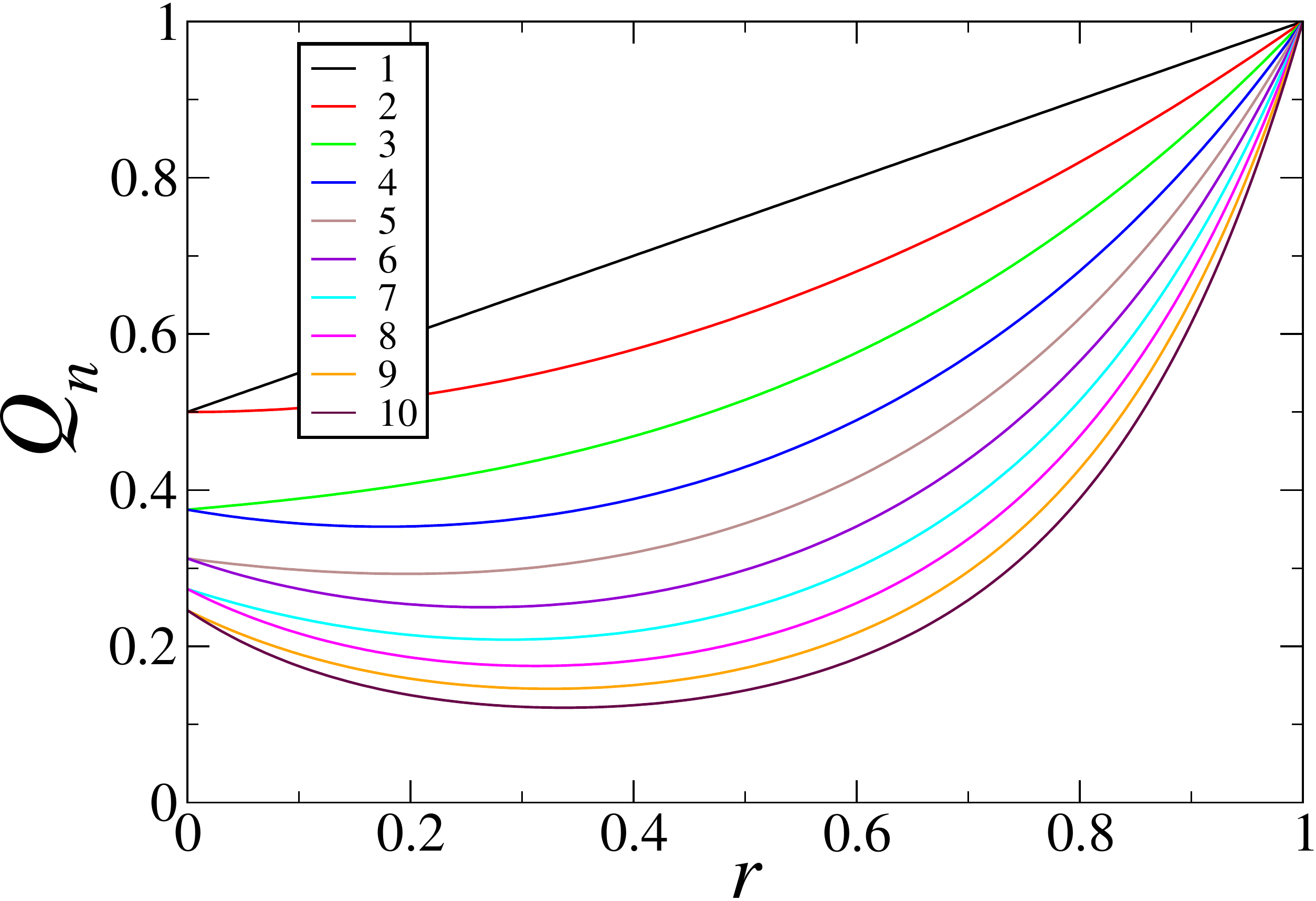}
\caption{\small
Survival probability $Q_n$ of the simple Polya walk
against resetting probability $r$ for times $n$ up to 10 (see legend).}
\label{qpolya}
\end{center}
\end{figure}

The survival probability $Q_n$ is plotted against~$r$ in figure~\ref{qpolya} up to $n=10$.
Its first values read
\beqa
Q_0=1,
\nonumber\\
Q_1=(1+r)R_0,\qquad Q_2=(1+r^2)R_0,
\nonumber\\
Q_3=(1+r)R_1,\qquad Q_4=(1+r^2)R_1,
\nonumber\\
Q_5=(1+r)R_2,\qquad Q_6=(1+r^2)R_2,
\nonumber\\
Q_7=(1+r)R_3,\qquad Q_8=(1+r^2)R_3,
\eeqa
with
\beqa
R_0=\frac{1}{2},
\nonumber\\
R_1=\frac{1}{8}(3-2r+3r^2),
\nonumber\\
R_2=\frac{1}{16}(5-8r+14r^2-8r^3+5r^4).
\nonumber\\
R_3=\frac{1}{128}(35-94r+205r^2-228r^3+205r^4-94r^5+35r^6).
\eeqa

The formula~(\ref{QR}) implies
\beq
(1+r)Q_{2k+2}-(1+r^2)Q_{2k+1}=0.
\eeq
This suggests to consider the same differences in the other parity sector,
namely
\beqa
\Delta_k
&=&(1+r)Q_{2k+3}-(1+r^2)Q_{2k+2}
\nonumber\\
&=&(1+r)^2R_{k+1}-(1+r^2)^2R_k.
\label{rdiff}
\eeqa
The corresponding generating series reads
\beq
\w\Delta(y)=(1+r)^2\frac{\w R(y)-R_0}{y}-(1+r^2)^2\w R(y).
\eeq
Using~(\ref{serR}), we obtain
\beq
\w\Delta(y)=\frac{1}{y^2}\left(\sqrt{1-(1-r)^2y}-1+\frac{1}{2}(1-r)^2y\right),
\eeq
hence
\beq
\Delta_k=-\frac{b_{k+2}}{2k+3}\,(1-r)^{2k+4},
\label{delta}
\eeq
(see~(\ref{bdef})).
The resulting two-term linear recursion for the polynomials $R_k$,
\beq
(1+r)^2R_{k+1}-(1+r^2)^2R_k=-\frac{b_{k+2}}{2k+3}\,(1-r)^{2k+4}\quad(k\ge0),
\label{rectwo}
\eeq
supersedes the three-term linear recursion
that could be derived by means of the algebraic approach of
section~\ref{contex}.

The behavior of $Q_n$ as $r\to0$
can be readily investigated by expanding the generating series~(\ref{serpolya}) as
\beqa
\w Q(z)
&=&\frac{1}{z}\left(\frac{1+z}{\sqrt{1-z^2}}-1\right)
\nonumber\\
&+&\frac{1}{z}\left(-\frac{1+z-z^2}{(1-z)\sqrt{1-z^2}}+\frac{1+z}{1-z}\right)r+\cdots,
\eeqa
hence
\beq
Q_n=q_n+c_n r+\cdots,\qquad c_n=2-(n+2)q_n.
\label{polya0}
\eeq

The survival probability $q_n$ in the absence of resetting
depends on the parity of~$n$ according to~(\ref{qnpolya}),
whereas the relation between $c_n$ and $q_n$ holds irrespective of
the parity of $n$.
The correction terms $c_1=1/2$ and $c_3=1/8$ are positive,
whereas~$c_2$ vanishes, and $c_4=-1/4$ and all subsequent ones are negative.
This corroborates the observation (see figure~\ref{qpolya})
that $Q_n$ is monotonically increasing with $r$ for $n=1$, 2 and~3,
whereas it exhibits a minimum for a non-trivial $r_n$ for all $n\ge4$.
The value $r_n$ of the resetting probability
at which $Q_n$ is minimal has been plotted against $1/n$ in figure~\ref{rn} (blue dataset).
Oscillations according to the parity of $n$ are clearly visible for the smaller values of $n$.
For large times, $r_n$ approaches the limit~(\ref{rmaxpolya}) (blue arrow),
at which the asymptotic decay rate $K$~(\ref{kpolya}) is maximal.

In the strong-resetting regime ($r\to1$), the result~(\ref{cont1}) holds unchanged,
as well as its interpretation.

\subsection{Asymptotic late-time results}
\label{polyasy}

The analysis of the asymptotic behavior of the survival probability $Q_n$ at
late times
closely follows the analysis made in section~\ref{contasy}.
For any fixed resetting probability $r$,
the closest singularity of the generating series~(\ref{serpolya})
is a simple pole located at
\beq
z_0=\frac{1+r}{1+r^2},
\eeq
whereas the branch-point singularity of the square root lies further away, at
\beq
z_c=\frac{1}{1-r}.
\eeq
The survival probability therefore falls off exponentially as
\beq
Q_n\approx A\,\e^{-Kn},
\label{exppolya}
\eeq
with
\beq
K=\ln z_0=\ln\frac{1+r}{1+r^2}
\label{kpolya}
\eeq
and
\beq
A=\frac{4r}{(1+r)^2}.
\label{Apolya}
\eeq
The expression~(\ref{kpolya}) of the decay rate $K$
can be alternatively derived from the recursion~(\ref{rectwo}).
This is however not the case for the amplitude $A$.

The decay rate $K$~(\ref{kpolya}) entering the exponential law~(\ref{exppolya})
has been plotted against~$r$ in figure~\ref{kplot} (blue curve).
It vanishes as $r\to0$ and $r\to1$, according to
\beq
\matrix{
K=r-\frad{3r^2}{2}+\cdots\hfill & (r\to0),\cr
K=\frad{1-r}{2}-\frad{(1-r)^2}{8}+\cdots\quad & (r\to1),
}
\label{kp01}
\eeq
and reaches its maximum,
\beq
K_{\rm max}=\ln\frac{1+\sqrt{2}}{2}\approx0.188226,
\label{kmaxpolya}
\eeq
for the value
\beq
r=\sqrt{2}-1\approx0.414213
\label{rmaxpolya}
\eeq
of the resetting probability.

Finally, a crossover similar to that described at the end of
section~\ref{contasy}
takes place in the scaling regime where $n$ is large, whereas $r$ is small.

\section{Arithmetic distributions}
\label{arith}

This section is devoted to the special class of arithmetic step length
distributions corresponding to random walks on the lattice of integers.
A study of the statistics of records for these distributions
in the absence of resetting is given in~\cite{mms} (see also~\cite{feller2}), while the case with resetting is addressed in~\cite{us}.

\subsection{General formalism and results}
\label{arithgal}

We focus our attention onto symmetric arithmetic step length distributions
with finite range $J$, of the form
\beq
\rho(\eta)=\sum_{j=1}^J\frac{a_j}{2}\left(\delta(\eta-j)+\delta(\eta+j)\right),
\qquad\sum_{j=1}^J a_j=1.
\label{rhoarith}
\eeq
For the sake of simplicity, steps of length zero are not allowed.
We consider the generic case where all the $a_j$ are non-zero.

The Fourier transform of~(\ref{rhoarith}) reads
\beq
\F\rho(k)=\sum_{j=1}^J a_j\cos(jk)=\sum_{j=1}^J a_jT_j(\cos k).
\label{fouarith}
\eeq
We have indeed $\cos(jk)=T_j(\cos k)$,
where the $T_j$ are the Tchebyshev polynomials of the first kind.
As a consequence, the Fourier transform $\F\rho(k)$
is a polynomial of degree~$J$ in $\cos k$.
It is therefore an even $2\pi$-periodic function of $k$.
In the $k\to0$ limit, we have
\beq
\F\rho(k)=1-Dk^2+\cdots,
\label{kzero}
\eeq
where
\beq
D=\frac{1}{2}\sum_{j=1}^J j^2a_j
\label{Ddef}
\eeq
is the diffusion coefficient, such that $\mean{\eta_n^2}=2D$.

The connection with the formalism of section~\ref{sa} goes as follows.
The probability introduced in~(\ref{Pdef}) reads
\beq
P_n=\int_0^\pi\F\rho(k)^n\,\frac{\dd k}{\pi},
\eeq
and therefore
\beq
\sum_{n\ge1}\frac{P_n}{n}\,z^n=-\int_0^\pi\ln(1-z\F\rho(k))\frac{\dd k}{\pi},
\eeq
so that~(\ref{gqser}) and~(\ref{Edef}) respectively become
\beq
\w
q(z)=\frac{1}{\sqrt{1-z}}\exp\left(-\frac12\int_0^\pi\ln(1-z\F\rho(k))\frac{\dd
k}{\pi}\right)
\label{gqarith}
\eeq
and
\beq
E=\exp\left(-\frac12\int_0^\pi\ln(1-\F\rho(k))\frac{\dd k}{\pi}\right).
\label{Earith}
\eeq
The last two formulas are given in~\cite{mms}.

Here, we propose to shed some new light on the problem by means of the
following algebraic approach.
Let us consider first the generating series $\w q(z)$, given
by~(\ref{gqarith}).
The expression~(\ref{fouarith}) of the Fourier transform $\F\rho(k)$
implies that $1-z\F\rho(k)$ is a polynomial of degree~$J$ in $\cos k$,
which can be factored as
\beq
1-z\F\rho(k)=C\prod_{j=1}^J(1-\Lambda_j\cos k),
\label{gqfactor}
\eeq
where the prefactor $C$ and the roots $\Lambda_j$ depend on $z$ and on the
weights $a_j$.
Setting $k=0$, we get
\beq
1-z=C\prod_{j=1}^J(1-\Lambda_j).
\label{gq0}
\eeq
Inserting~(\ref{gqfactor}) into~(\ref{gqarith}),
and using the integral
\beq
\int_0^\pi\ln(1-\Lambda\cos k)\frac{\dd k}{\pi}
=-2\ln\frac{\sqrt{1+\Lambda}-\sqrt{1-\Lambda}}{\Lambda},
\label{logint}
\eeq
as well as the identity~(\ref{gq0}),
we obtain the expression
\beq
\w
q(z)=\frac{1}{C}
\prod_{j=1}^J\frac{1}{\Lambda_j}\left(\sqrt{\frac{1+\Lambda_j}{1-\Lambda_j}}-1\right).
\label{gqprod}
\eeq

The generating series $\w q(z)$ is an algebraic function of $z$ and
the $a_j$ with degree~$2^J$.
The right-hand side of~(\ref{gqprod}) is indeed a symmetric function of the roots~$\Lambda_j$,
where each square root has two branches,
and therefore brings a factor two to the degree of~$\w q(z)$.

The above property still holds in the presence of resetting.
The key relation~(\ref{sres}) between $\w q(z)$ and $\w Q(z)$ is rational,
and so the generating series $\w Q(z)$
is an algebraic function of $z$, $r$ and the $a_j$, with the same degree~$2^J$.
As a consequence, $Q_n$ obeys linear recursion relations,
generalizing~(\ref{recfour}), whose complexity grows exponentially
with the range $J$ of the walk.
Furthermore, the location $z_0$ of the closest singularity of the generating
series~$\w Q(z)$,
governing the exponential decay of $Q_n$,
is an algebraic function of the resetting probability $r$ and the $a_j$,
whose degree is however not determined by the present reasoning.

Consider now the asymptotic enhancement factor $E$, given by~(\ref{Earith}).
The difference $1-\F\rho(k)$ vanishes for $k=0$,
and can therefore be factored as
\beq
1-\F\rho(k)=c(1-\cos k)\prod_{j=1}^{J-1}(1-\lambda_j\cos k).
\label{dprod}
\eeq
where the prefactor $c$ and the roots $\lambda_j$ depend on the weights
$a_j$.
In the $k\to0$ limit, using~(\ref{kzero}),~(\ref{Ddef}), we obtain
\beq
2D=c\prod_{j=1}^{J-1}(1-\lambda_j).
\label{d0}
\eeq
Inserting~(\ref{dprod}) into~(\ref{Earith}),
and using~(\ref{logint}) and~(\ref{d0}),
we obtain
\beq
E=\frac{1}{\sqrt{D}}
\prod_{j=1}^{J-1}\frac{1-\lambda_j}{\lambda_j}\left(\sqrt{\frac{1+\lambda_j}{1-\lambda_j}}-1\right).
\label{eprod}
\eeq
The product
\beq
\mean{h_1}=E\sqrt{D}
\label{spitz}
\eeq
is therefore an algebraic function of the $a_j$ with degree $2^{J-1}$.
This quantity is the mean value of the first positive abscissa $h_1$
of a random walk issued from the origin,
irrespective of the time at which this position is
reached~\cite{spitzer3,spitzerbook,us}
(see also~\cite[ch.~XVIII]{feller2}).

\subsection{The simple Polya walk}

A first illustration of the above approach is provided by the simple Polya
walk,
studied at length in section~\ref{polya}.
In this case, we have $J=1$, $a_1=1$, $D=1/2$ and $\w\rho(k)=\cos k$.
For generic values of $z$, we have $C=1$ and $\Lambda_1=z$,
and so~(\ref{gqprod}) reads
\beq
\w q(z)=\frac{1}{z}\left(\sqrt{\frac{1+z}{1-z}}-1\right),
\eeq
which is equivalent to~(\ref{qserpolya}).
Moreover,~(\ref{eprod}) yields $\mean{h_1}=E\sqrt{D}=1$, as should be.

\subsection{Lattice walks with range two}

The next case, in order of increasing complexity,
consists in lattice walks with range $J=2$.
Using the parametrization
\beq
a_1=1-a,\qquad a_2=a,
\eeq
we have
\beq
D=\frac{1+3a}{2}
\eeq
and
\beq
\w\rho(k)=(1-a)\cos k+a\cos(2k)=(1-a)\cos k+a(2\cos^2k-1).
\eeq
For generic values of $z$, we have therefore
\beq
1-z\w\rho(k)=1+az-(1-a)z\cos k-2az\cos^2k,
\eeq
so that $C=1+az$, whereas $\Lambda_1$ and $\Lambda_2$ are the roots of the quadratic
equation
\beq
(1+az)\Lambda^2-(1-a)z\Lambda-2az=0.
\label{quad}
\eeq
By eliminating the roots $\Lambda_1$ and $\Lambda_2$ between~(\ref{gqprod})
and~(\ref{quad}), we obtain
\beqa
a^2z^2(1-z)^2\w q(z)^4+2az(1-z)^2\w q(z)^3
\nonumber\\
-2(1+a)z(1-z)\w q(z)^2-4(1-z)\w q(z)+4=0.
\label{qfour}
\eeqa
As anticipated, this is a polynomial equation with degree $4$ for $\w q(z)$,
the generating series for the survival probability $q_n$ in the absence of
resetting.

As far as the enhancement factor $E$ is concerned,
the constants entering the product~(\ref{dprod}) read $c=1+a$ and
$\lambda_1=-2a/(1+a)$,
and so~(\ref{eprod}) yields
\beq
E=\frac{\sqrt{1+3a}-\sqrt{1-a}}{a\sqrt{2}}.
\label{eres}
\eeq
The enhancement factor reaches its maximum $E_{\rm max}=\sqrt2$,
corresponding to the simple Polya walk (see~(\ref{epolya})),
for $a=0$ and $a=1$,
and its minimum $E_{\rm min}=\sqrt{3/2}$ for $a=2/3$.
The product
\beq
\mean{h_1}=E\sqrt{D}=\frac{2}{1+\sqrt\frad{1-a}{1+3a}}
\eeq
increases monotonically from 1 to 2 as $a$ increases from 0 to 1.

In the presence of resetting, using the key relation~(\ref{sres}),
we can derive a polynomial equation with degree $4$ for $\w Q(z)$,
that is similar to~(\ref{qfour}), albeit too long to be reported here.
The location $z_0=\e^K$ of the closest pole of $\w Q(z)$,
which governs the exponential decay $Q_n\sim\e^{-Kn}$,
is found to obey the quadratic equation
\beqa
\left((1-r)^4a^2+2r(1-r)^2a+2r^2(1+r^2)\right)z_0^2
\nonumber\\
-2\left((1-r)^3a^2+r(1-r)(2-r)a+r^2(1+r)\right)z_0
\nonumber\\
+(1-r)^2a^2+2r(1-r)a=0.
\eeqa
The corresponding decay rate $K$ vanishes as $r\to0$ and as $r\to1$, as
\beq
\matrix{
K=r+\left(\frad{1}{2}-E^2\right)r^2+\cdots\hfill & (r\to0),\cr
K=\frad{1-r}{2}-\frad{1-2a(1-a)}{8}(1-r)^2+\cdots\quad & (r\to1).
}
\eeq
The first expression involves the enhancement factor $E$ given in~(\ref{eres}).
Both expansions agree with~(\ref{kp01}) for $a=0$ and $a=1$, as should be.

\section{Discussion}
\label{disc}

In a recent work~\cite{us},
we have revisited various features
of the statistics of extremes and records of symmetric random walks
with stochastic resetting.
In the present paper, which is a sequel to the latter,
we entirely focus our attention onto
the survival probability (or persistence probability) $Q_n$,
defined as the probability for the walker not to cross the origin
up to time $n$.
Throughout this work we consider for simplicity symmetric step length distributions,
and the origin is both the starting point of the walker and its resetting point.

Stochastic resetting with probability $r$ at each step
has the peculiar feature that it does not alter the Markovian nature of free random walk.
This entails, among other remarkable properties,
the existence of the identity~(\ref{sresintro}) (or~(\ref{sres}))
between the generating series of the survival probabilities~$Q_n$ and~$q_n$, respectively
with and without resetting.
In the present situation,
this identity allowed us to investigate in detail
many facets of the problem at hand.

For random walks and L\'evy flights with symmetric and continuous
step length distributions,
the survival probabilities are universal:
$q_n$ only depends on time $n$ according to~(\ref{qn}),
whereas $Q_n$ only depends on $n$ and $r$.
The properties of $Q_n$ are explored at depth in section~\ref{cont},
both at finite times and in the asymptotic regime of late times.
Among their most noticeable properties,
the $Q_n$ are polynomials of degree~$n$ in $r$ which obey a four-term linear recursion,
as a consequence of the algebraic nature of the generating series $\w Q(z)$.
They fall off exponentially in~$n$,
with the corresponding decay rate $K$ vanishing both as $r\to0$ and as $r\to1$,
and being maximal for $r=2/5$.

Whenever the step length distribution is not continuous,
the survival probabilities are no longer universal,
but rather depend on details of the underlying distribution.
The class of arithmetic step length distributions, that is,
discrete probability distributions yielding random walks on the lattice of integers,
deserves special interest.
The simplest of all lattice walks is the simple Polya walk,
for which a thorough investigation of the
survival probability $Q_n$ is given in section~\ref{polya},
both at finite times and in the asymptotic regime of late times.
The $Q_n$ are again polynomials of degree~$n$ in~$r$,
whose structure depends on the parity of $n$.
They fall off exponentially in~$n$,
with the corresponding decay rate $K$ being maximal for $r=\sqrt{2}-1$.
In the generic situation of lattice walks with a larger range,
we have shown that quantities such as the generating series $\w Q(z)$
and the asymptotic enhancement factor $E$ are algebraic functions
of the model parameters, and we determined their respective degrees.
Algebraic functions were shown to play a key role in a germane problem,
namely the encounters of the simple Polya walk with a ballistic obstacle~\cite{bgl}.
The algebraic approach can be expected to facilitate further investigations of lattice walks.

\section*{References}

\bibliography{paper.bib}

\end{document}